\documentclass{article}

\usepackage[paperwidth=8.5in]{geometry}

\usepackage{url}
\usepackage{graphicx}
\usepackage{hyperref}
\usepackage{authblk}

\usepackage{pdfpages}

\newcommand{\TRONCO}{\texttt{TRONCO}}
\newcommand{\R}{\textsc{R}}

\newcommand{\rev}[1]{\textcolor{black}{#1}}

\begin{document}

\title{\TRONCO{}: an \R{} package for the inference of cancer progression models from heterogeneous genomic data}

\author[1]{Luca De Sano\footnote{Equal contributors}$\,$,}
\author[1,2]{Giulio Caravagna$^{\ast}$,}
\author[1]{Daniele Ramazzotti$^{\ast}$,}
\author[1,3]{Alex Graudenzi}
\author[1]{Giancarlo Mauri}
\author[4]{Bud Mishra}
\author[1,5]{Marco Antoniotti}

\affil[1]{Department of Informatics, Systems and Communication, University of Milano-Bicocca, Milan, Italy.}
\affil[2]{School of Informatics, University of Edinburgh, Edinburgh, UK.}
\affil[3]{Institute of Molecular Bioimaging and Physiology of the Italian National Research Council (IBFM-CNR), Milan, Italy.}
\affil[4]{Courant Institute of Mathematical Sciences, New York University, New York, USA.}
\affil[5]{Milan Center for Neuroscience, University of Milan-Bicocca, Italy.}

\date{}

\maketitle

\begin{abstract}

  \textbf{Motivation:} We introduce \TRONCO{} (TRanslational
  ONCOlogy), an open-source \R{} package that implements the
  state-of-the-art algorithms for the inference of cancer progression
  models from (epi)genomic mutational profiles. \TRONCO{} can be used
  to extract population-level models describing the trends of
  accumulation of alterations in a cohort of cross-sectional samples,
  e.g., retrieved from publicly available databases, and
  individual-level models that reveal the clonal evolutionary history
  in single cancer patients, when multiple samples, e.g., multiple
  biopsies or single-cell sequencing data, are available.  The
  resulting models can provide key hints in uncovering the
  evolutionary trajectories of cancer, especially for precision
  medicine or personalized therapy.

  \textbf{Availability:} \TRONCO{} is released under the GPL license,
  it is hosted in the Software section at
  \href{http://bimib.disco.unimib.it/}{http://bimib.disco.unimib.it/}
  and archived also  at
  \href{http://www.bioconductor.org/}{bioconductor.org}.

  % and \href{https://github.com/BIMIB-DISCo/TRONCO}{https://github.com/BIMIB-DISCo/TRONCO}. 

  \textbf{Contact:} \href{tronco@disco.unimib.it}{tronco@disco.unimib.it} 

\end{abstract}

\section{Introduction}
\label{sec:introduction}

\begin{figure*}[!ht]
  \begin{center}
    \includegraphics[width=1.00\textwidth]{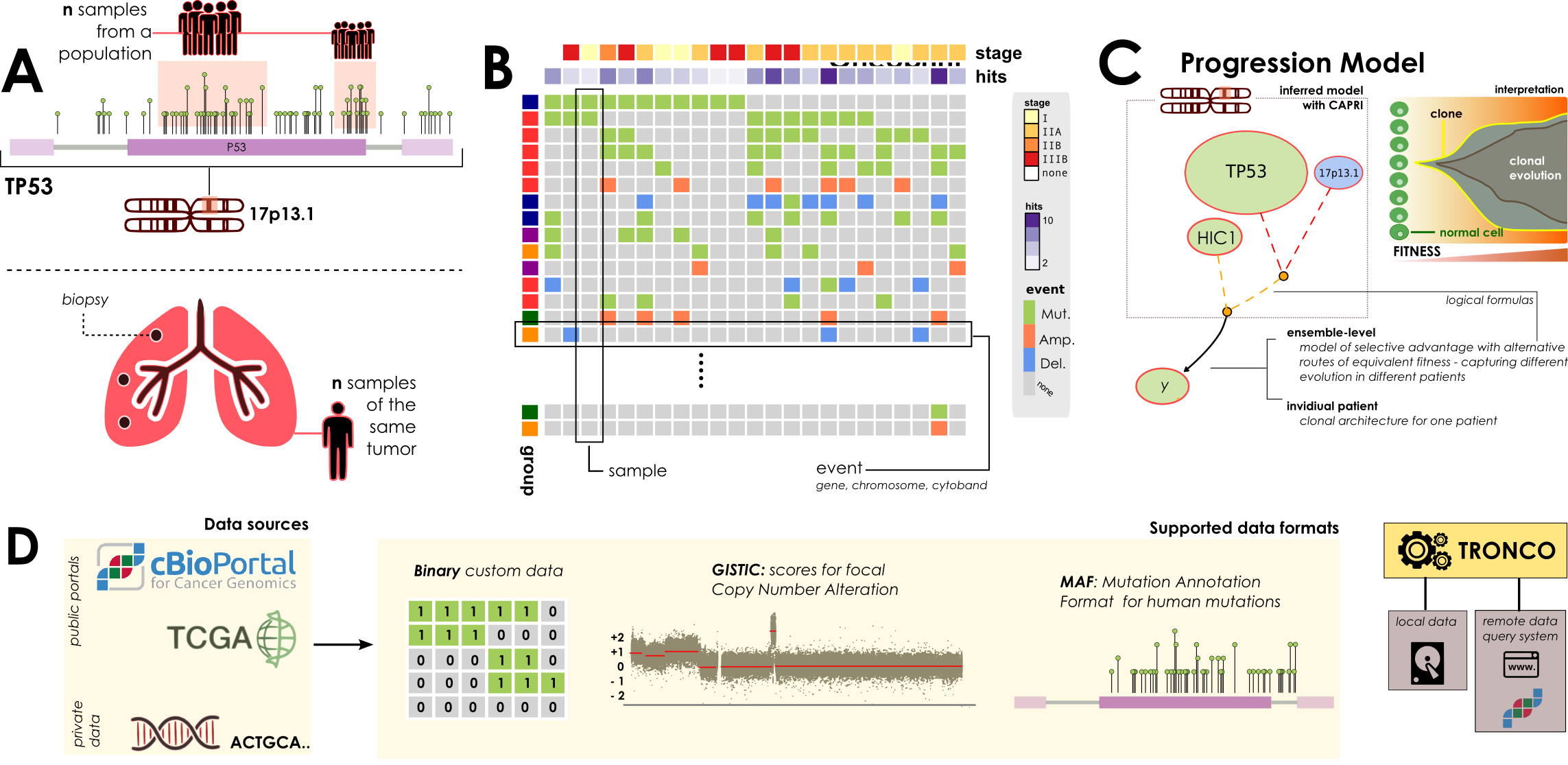}
  \end{center}
  \vspace*{.05in}
  \caption{\textbf{(A)} \texttt{TRONCO} can process either alterations
    (e.g., somatic mutations or wider chromosomal lesions) in a cohort
    of independent samples (top lolliplot diagram), or a set of
    multiple snapshots from a unique patient (e.g., multi-region or
    single-cell, bottom panel).  \textbf{(B)} \emph{Oncoprints} allow the
    user to visualize the data that the tool is processing.
    Regardless of the source, each row represents a certain
    alteration - at a custom resolution depending on the cancer under study  - and each column a sample. \textbf{(C)} A model inferred
    with the tool might outline  cancer evolution occurring in a population 
     ensemble or in an individual patient. \rev{Graphically, alterations are
     represented as nodes with different colors (e.g., green mutations
     and blue homozygous deletions). Algorithms such as CAPRI 
     allow describing  alterations with logical formulas, in an attempt to
     find their role as a ``group''  (see  \cite{Ramazzotti15092015}
     for details); we picture such groups  with dashed lines.} In the panel, we show a
    hypothetical ensemble-level model predicting a selection pressure
    on two genes mapped to  17p13, \textsc{tp53} and \textsc{hic1}, as
    it may be inferred by analyzing  samples  harbouring either
    \textsc{tp53}/\textsc{hic1} mutations or homozygous deletions in
    the cytoband where any of these two genes map, i.e., here for
    purely explanatory cases we suppose just \textsc{tp53}, which maps
    to 17p13.1. The model suggests a trend of selection  towards
    mutations in gene \textsc{y}, which shall be interpreted as a  set
    of preferential  clonal expansions characteristic of the population of analyzed samples,  involving
    alterations of the functions mapped to 17p13 and \textsc{y}.
    \rev{\textbf{(D)} \texttt{TRONCO} supports  three data types. Custom data, which is supposed
    to be provided as a {\em binary input matrix} storing the presence (1) or absence (0) of a certain alteration
    in a sample. Or, standard data formats such as the \emph{Mutation Annotation
   Format} (MAF) for  somatic mutations, as well as the \emph{Genomic
   Identification of Significant Targets in Cancer}  (GISTIC) format for focal Copy Number Variations.
   Data can be generated by custom experiments, or collected - along with other ``omics'' - 
   from public databases such as TCGA and cBio portal. For the latter, cBio portal, \texttt{TRONCO} implements a query
    system to fetch data with minimal effort. The tool engine can then be used to manipulate  genomic
    profiles -- regardless of their source -- and  run progression inference  algorithms.
	}
    }
\label{fig:XXX}
\end{figure*}

Cancer develops through the successive expansions of clones, in which 
certain (epi)genomic alterations, called \emph{drivers}, confer a
\emph{fitness} advantage and progressively accumulate, in a context of
overall scarcity of resources \cite{nowell1976clonal}. \rev{Specifically, in Nowell's seminal work, tumor evolution is described in terms of stepwise genetic variation such that growth advantage is the key for the survival and prorogation of the clones.} Therefore, one can define \emph{cancer progression models}, in terms of
\emph{probabilistic causal graphical models}, where the conditional
dependencies and the temporal ordering among these alterations are described, revealing the evolutionary trajectories of cancer at the
(epi)genome level.

We further distinguish \cite{Caravagna2015:_pipeline}. $(i)$
\emph{ensemble-level} progression models, describing the statistical
trends of accumulation of genomic alterations in a cohort of distinct
cancer patients. \rev{Such models describe the temporal partial orders of fixation and accumulation of such alterations and represent population-level trends;} and $(ii)$  \emph{individual-level} models, thus
accounting for the specific evolutionary history of cancer clones in
individual tumors. \rev{Such models thus impute the ancestry relations of the observed clones.} 

Even if the inference of such models is further complicated by a
series of theoretical and technical hurdles, such as, e.g.,
\emph{intra-} and \emph{inter-tumor heterogeneity} and the effective
detection of drivers, it can benefit from the increasing amount of
\emph{next-generation sequencing} (NGS) data, currently available
through public projects such as The Cancer Genome Atlas (TCGA,
\url{https://tcga-data.nci.nih.gov}).
%\cite{tcga}.
Usually, such databases provide \emph{cross-sectional} (epi)genomic
profiles retrieved from single biopsies of cancer patients, which can
be used to extract ensemble-level models; but higher resolution data
such as \emph{multiple-biopsies}, or even \emph{single-cell} sequencing
data are becoming more accessible and reliable, which
can be used to infer individual-level models.

Here we introduce \TRONCO{} (TRanslational ONCOlogy), an \R{} package
built to infer cancer progression models from heterogeneous genomic
data (in the form of alterations persistently present along tumor
evolution.)  Currently, \TRONCO{}  provides the implementation of two 
algorithms: $(i)$  \texttt{CAPRESE} (CAncer PRogression
Extraction with Single Edges \cite{Loohuis:2014im}), and $(ii)$
\texttt{CAPRI} (CAncer PRogression Inference
\cite{Ramazzotti15092015}), both based on Suppes' theory of
\emph{probabilistic causation} \cite{suppes1970probabilistic}, but
with distinct goals and properties (see  Software Implementation).

\TRONCO{}, in its current form and perspective, should be thought
of as a tool that provides the implementation of up-to-date solutions
to the progression inference problem.  At the time of the writing it 
can be effectively used as the final stage of a modular
pipeline for the extraction of ensemble-level cancer progression
models from cross-sectional data \cite{Caravagna2015:_pipeline}. In
such a pipeline input data are pre-processed to $(i)$ stratify samples
in tumor subtypes, $(ii)$ select driver alterations and $(iii)$
identify groups of fitness-equivalent (i.e., mutually exclusive)
alterations, prior to the application of the \texttt{CAPRI} algorithm.
The resulting ensemble-level progression models depict the evolutionary dynamics of cancer, with 
translational impacts on diagnostic and therapeutic processes, especially in
regard to precision medicine and personalized drug development. 

From the complementary perspective, \texttt{TRONCO} can also exploit
the \texttt{CAPRESE} algorithm to infer the clonal evolutionary
history in single patients when multiple samples are available, as in
the case of multiple biopsies and/or single-cell sequencing data, as
long as the set of driver events is selected; see
\cite{Caravagna2015:_pipeline}.

\section{Software Implementation}
\label{sec:algorithms}

\TRONCO{} implements a set of \R{} functions to aid the user  to
extract a cancer progression model from genomic data. At a high-level,
these function shall help to import, visualize and manipulate  genomic
profiles -- regardless of their source -- eventually allowing
the implemented algorithms to run and assess the
confidence in a model.

\rev{The basics steps of \TRONCO{}'s usage are shown in Figure \ref{fig:XXX}. 
In panel $(a)$ we show multiple  input  alterations (e.g., somatic mutations or copy number alterations) either from a cohort of patients, or a unique patient (e.g., multi-region or single-cell sequencing); in panel $(b)$ we show an oncoprint visualization from the tool, i.e., a matrix whose columns represent  samples and rows the  alterations and their presence per sample; panel $(c)$ shows an inferred  graphical Bayesian progression model    obtained with one of the available algorithms; finally, in panel $(d)$ we show data supported for processing in the tool. For a more detailed explanation of the implementation of the package see \cite{antoniotti2015design}.}

% descrivere oncoprint, descrivere provenienza dati, togliere pipeline, indicare che i modelli sono grafi

\paragraph{Data loading and manipulation.} Common  formats to store
data used to extract progression models can be natively imported. 
These include, for instance, the  \emph{Mutation Annotation
  Format} (MAF) for  somatic mutations, as well as the \emph{Genomic
  Identification of Significant Targets in Cancer}  (GISTIC) format to
store focal Copy Number Variations. The tool can exploit the cBio
portal for Cancer Genomics, which collects among others TCGA projects,
to access freely available instances of such data
\cite{cerami2012cbio}.

% ref al pannello D quando ci sarà

\TRONCO{} provides functions for data preprocessing to, e.g., select a certain subset of alterations, or samples
or any abstraction which might be appropriate according to the cancer being studied.

\paragraph{Visualization and interaction with other tools.} \TRONCO{}
implements an \emph{oncoprint} system to visualize the processed data. Datasets can be
exported for processing by other tools used to,
e.g., stratify input samples and detect groups of mutually exclusive
alterations, which include the \emph{Network Based Stratification}
\cite{hofree2013network} and \emph{MUTEX} \cite{babur2015systematic}
tools. \TRONCO{} allows the visualization of the inferred models.

\paragraph{Model inference and confidence.} \TRONCO{} provides two 
algorithms: $(i)$ \texttt{CAPRESE}, which uses a
\emph{shrinkage}-like estimator to infer \emph{tree}-models of
progression, and $(ii)$ \texttt{CAPRI}, which extracts more general
\emph{direct acyclic graphs (DAG)} - thus allowing for confluent
evolution and complex hypothesis testing -- by combining
\emph{bootstrap} and \emph{maximum likelihood estimation}. \rev{\texttt{CAPRESE} and \texttt{CAPRI} both rely on the same theory of \emph{probabilistic causation}, but with distinct goals and properties. The former reconstructs tree models of  progressions, while the latter  general directed acyclic graphs. Both methods are agnostic  to the type of input data (i.e., whether its an ensemble or an individual tumor), but shall be used in different contexts as they produce different types of models. Indeed, \texttt{CAPRESE} is better at extracting cancer evolution in a single individual as in that case trees capture branched evolution and trunk events, which shall suffice to describe clonal evolution. Instead, when heterogeneity might result in multiple evolutionary routes with common downstream alterations, the underlying true model is a graph, and \texttt{CAPRI} should be the tool of choice.} %In \cite{Caravagna2015:_pipeline} examples of the usage of the two algorithms on real data along with a deeper discussion on the interpratation of the models are provided.} 

% togliere ref a suppes / caprese si presta a ricostruire architetture clonali su singolo pz - capri visto che permette di...  togliere ref a caravagna&

Whatever a model is, \texttt{TRONCO} implements a set of functions to assess
its confidence   via $(i)$ non-parametric,
$(ii)$ parametric and $(iii)$ statistical bootstrap.

\section{Discussion}
\label{sec:discussion}

% \TRONCO{} aims to become the bioinformatics reference tool of choice
% providing  the  community  with the implementation of up-to-date
% statistical routines to understand the evolutionary trajectories of
% any cancer. Indeed, in terms of computational speed, scalability with
% respect to sample size, predictive accuracy and robustness against
% noise in the input data, the algorithms implemented in the tool is
% demonstrably to be the current state-of-the-art for the inference
% problem.  The implementation in \R{} makes straightforward the
% interaction of \TRONCO{} with other common bioinformatics tools,
% possibly allowing the creation of a common suite of tools for cancer
% progression inference. The porting of the tool to other widely used
% bioinformatics platforms such as, e.g., Cytoscape
% \cite{shannon2003cytoscape} is under development.

\TRONCO{} provides up-to-date, theoretically well-founded, statistical
methods to understand the evolution of a cancer
(ensamble-level) or a single tumor (individual-level).
The implemented algorithms are demonstrably
the state-of-the-art for the progression inference problem, in
terms of computational cost, scalability with respect to sample size, 
accuracy and robustness against noise in the data.
The implementation makes straightforward the interaction of
\TRONCO{} with other common bioinformatics tools, possibly allowing
the creation of a common suite of tools for cancer progression
inference.

\rev{Finally, we refer to \cite{Caravagna2015:_pipeline} or the Supplementaty materials for a demonstration
of the usage of \TRONCO{} on real genomics data both at the ensemble-level
and individual-level progression models. In particular, this paper outlines the capability of the methods to reproduce much of the current knowledge
on the progression for a set of cancer types, as well as to suggest clinically relevant 
insights. Furthermore, we also provide users with detailed manuals, vignettes, and source code to replicate all the analysis presented in the paper plus others (case studies: colorectal cancer, clear cell renal cell carcinoma and acute chronic myeloid leukaemia) in the Supplementary Materials and at the \TRONCO{} official webpage (Software section at \href{http://bimib.disco.unimib.it/}{http://bimib.disco.unimib.it/}).}
 
\paragraph{\textbf{Financial support.}} MA, GM, GC, AG, DR acknowledge
Regione Lombardia (Italy) for the research projects RetroNet through
the ASTIL Program [12-4-5148000-40]; U.A 053 and Network Enabled Drug
Design project [ID14546A Rif SAL-7], Fondo Accordi Istituzionali 2009.
BM acknowledges founding  by the NSF grants CCF-0836649, CCF-0926166
and a NCI-PSOC grant.

\bibliographystyle{plain}
\bibliography{biblio}

% add the supplementary


\section*{Supplementary material}

# TRONCO: an R package for the inference of cancer progression models from heterogeneous genomic data

*Supplementary Information*

Luca De Sano    Giulio Caravagna    Daniele Ramazzotti
Alex Graudenzi    Giancarlo Mauri    Bud Mishra
Marco Antoniotti

# Contents

# 1 Installation

TRONCO's current stable release is version 2.0 (Mantish Shrimp, August 2015).

This can be installed from our GitHub account by typing within R the following commands:

```
> library(devtools)
> install_github("BIMIB-DISCo/TRONCO")
```

Mantis Shrimp is also in Bioconductor 3.2 since 14 October 2015.

# 2 Examples

TRONCO official webpage (http://bimib.disco.unimib.it/) hosts detail commentary and source code to replicate the following studies:

1. Atypical Chronic Myeloid Leukemia (aCML) - ensemble level;

2. Colorectal Cancer (CRC) - ensemble level;

3. Clear Cell Renal Cell Carcinoma (CCRCC) - individual level;

4. Single-cell synthetic data test for reconstruction's performance;

5. Processing data from the cBio portal (next section)

In the next subsections we present a few examples of usage of the tool.

## 2.1 Fetching data from the cBio portal

We show how to download *lung cancer data* of somatic mutations reported in **Figure 1(a)** of the paper

> Comprehensive molecular profiling of lung adenocarcinoma. *The TCGA Consortium.* Nature 511, 543-550, 2014.

from the archives available at cBio portal under url

http://www.cbioportal.org/study.do?cancer_study_id=luad_tcga_pub

The code that we show is tested under R's most recent version.

**Fetching data.** First, define the pool of genes for which one wishes to download data. In this case we have a list of 18 genes from the reference figure.

```
> genes = c('TP53', 'KRAS', 'KEAP11', 'STK11', 'EGFR', 'NF1',
>     'BRAF', 'SETD2', 'RBM10', 'MGA', 'MET', 'ARID1A', 'PIK3CA',
>     'SMARCA4', 'RB1', 'CDKN2A', 'U2AF1', 'RIT1')
```

Data in cBio portal are identified by the following triple:

- a study ID - a high-level references of the study, here `luad_tcga_pub`;

- a dataset ID - a set of samples which are part of the study, in our case `luad_tcga_pub_cnaseq`;

- a genetic profile ID - a type of data available for those samples, here `luad_tcga_pub_mutations`.

With these references, one can extract data from the portal. If these are not known in advance, one can run TRONCO's function `cbio.query` which wraps the CGDS-R package - the official query system for the portal, in R.

This function will help you by showing the available data at the portal:

```
> data = cbio.query(genes=genes)

*** CGDS plugin for cBio portal query.
Available studies at cBio portal.
```

```
        cancer_study_id   name
[ ... ]
45      luad_broad        Lung Adenocarcinoma (Broad, Cell 2012)
46      luad_tcga_pub     Lung Adenocarcinoma (TCGA, Nature 2014)
47      luad_tcga         Lung Adenocarcinoma (TCGA, Provisional)
[ ... ]
```

`> Enter cBio study id: luad_tcga_pub`

Here from a list of IDs archived at the portal, and we have selected the one which references the data used in the manuscript. Notice that the user can also access provisional TCGA data or datasets from other studies such as those carried out at the Broad Institute of MIT and Harvard.

The function `cbio.query` will then load dataset information and will allow the user to select the set of desired samples, in this case we select all tumor samples that have CNA and sequencing data.

`> Cancer codename: luad_tcga_pub`
```
Cancer Ref.: Lung Adenocarcinoma (TCGA, Nature 2014)
Cancer Syn.: TCGA Lung Adenocarcinoma, containing 230 samples;
Available datasets for study: luad_tcga_pub

      case_list_id                  case_list_description
1     luad_tcga_pub_3way_complete   All tumor samples that have
                                    mRNA, CNA and sequencing data
                                    (230 samples)
[ ... ]
22    luad_tcga_pub_cnaseq          All tumor samples that have CNA
                                    and sequencing data
                                    (230 samples)
```

`> Enter study dataset id: luad_tcga_pub_cnaseq`

Which opens a view over the available genetic profiles for the selected 230 samples.

```
Data codename: luad_tcga_pub_cnaseq
Data Ref.: Tumors with sequencing and CNA data
Data Syn.: All tumor samples that have CNA and sequencing data
           (230 samples)
Available genetic profiles for selected datasets.

      genetic_profile_id            genetic_profile_description
1     luad_tcga_pub_rna_seq_v2      mRNA z-Scores ...
      mrna_median_Zscores
[ ... ]
4     luad_tcga_pub_gistic          Putative copy-number calls on
```

```
                                        230 cases GISTIC 2.0.
[ ... ]
6    luad_tcga_pub_methylation_hm27 Methylation (HM27) beta-val ...
8    luad_tcga_pub_mutations        Mutation data from WES.
[ ... ]
11   luad_tcga_pub_mrna             Expression levels for 17235 ...
```

```
> Enter genetic profile id: luad_tcga_pub_mutations
```

If we wish to select other type of data for these samples we just re-run this script and select, e.g., mRNA z-Scores, expression values or Methylation beta-values.

```
Samples codename: luad_tcga_pub_mutations
Data Ref.: Mutations
Data Syn.: Mutation data from whole exome sequencing.

Querying the following list of genes: TP53, KRAS, ......, RIT1
Symbol "." was replaced with "-" in sample IDs.

Data retrieved: 230 samples, 17 genes.
Retrieved also clinical data for samples: luad_tcga_pub_cnaseq
Data exported to file: luad_tcga_pub.luad_tcga_pub_cnaseq.luad_tcga
```

In this way we collected 230 samples and 5 genes, plus the function automatically downloads clinical data for the selected samples. All these data are exported to an Rdata file named for this query

luad_tcga_pub.luad_tcga_pub_cnaseq.luad_tcga_pub_mutations.Rdata

so that further processing can happen offline.

**From data to a TRONCO object.** The returned object, here `data`, has flags to access the retrieved data. Mutation data is available through $profile tag, clinical data through $clinical tag.

```
> head(data$profile[, 1:5])
                ARID1A BRAF     CDKN2A   EGFR        KRAS
TCGA-05-4249-01 NaN    A762E    NaN      NaN         G12C
TCGA-05-4382-01 E1760* L613F    NaN      R222L,E545Q      NaN
TCGA-05-4384-01 NaN    NaN      NaN      NaN         NaN
TCGA-05-4389-01 NaN    G469V    NaN      NaN         NaN
TCGA-05-4390-01 NaN    NaN      NaN      NaN         G12V
TCGA-05-4395-01 NaN    NaN      NaN      NaN         G12V

> head(data$clinical[, 'HISTOLOGICAL_SUBTYPE', drop = F])
                HISTOLOGICAL_SUBTYPE
TCGA-55-7573-01 Lung Adenocarcinoma- Not Otherwise Specified (NOS)
```

4

```
TCGA-78-7150-01 Lung Adenocarcinoma- Not Otherwise Specified (NOS)
TCGA-55-7727-01 Lung Adenocarcinoma- Not Otherwise Specified (NOS)
TCGA-49-6767-01 Lung Adenocarcinoma- Not Otherwise Specified (NOS)
TCGA-49-6744-01 Lung Adenocarcinoma Mixed Subtype
TCGA-49-4488-01 Lung Adenocarcinoma- Not Otherwise Specified (NOS)
```

Basic conversion to this format to a representation which can be imported with TRONCO is straightforward, as we can go through a conversion to R's matrix, replace string values with corresponding 0s and 1s - as in this case we are not distinguishing types of mutations - , and transform the matrix in a tronco object via `import.genotypes`.

The imported dataset can be seen in the console with `show`, and visualized via `oncoprint`.

```
> m = as.matrix(data$profile)
> m[is.na(m)] = 0
> m[m == 'NaN'] = 0
> m[m != '0'] = 1
> tronco.data = import.genotypes(m, event.type = 'Mutation',
> color = 'brown3')
> tronco.data = annotate.description(tronco.data,
> "Lung cancer data from cBio portal")

> show(tronco.data, 5)
Description: Lung cancer data from cBio portal.
Dataset: n=230, m=17, |G|=17.
Events (types): Mutation.
Colors (plot): brown3.
Events (10 shown):
        G1 : Mutation ARID1A
        G2 : Mutation BRAF
        G3 : Mutation CDKN2A
        G4 : Mutation EGFR
        G5 : Mutation KRAS
Genotypes (10 shown):
                G1  G2  G3  G4  G5
TCGA-05-4249-01  0   1   0   0   1
TCGA-05-4382-01  1   1   0   1   0
TCGA-05-4384-01  0   0   0   0   0
TCGA-05-4389-01  0   1   0   0   0
TCGA-05-4390-01  0   0   0   0   1
TCGA-05-4395-01  0   0   0   0   1

> oncoprint(tronco.data)
*** Oncoprint for "Lung cancer data from cBio portal"
with attributes: stage=FALSE, hits=TRUE
```

```
Sorting samples ordering to enhance exclusivity patterns.
Setting automatic row font (exponential scaling): 10.7
```

The image obtained is shown in figure 1.

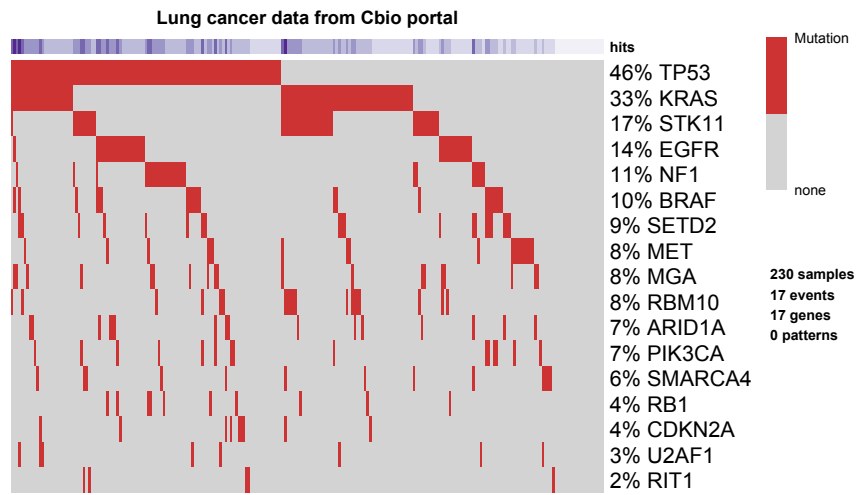

Figure 1: Oncoprint showing the genomic profiles obtained by quering cBio portal.

The reconstruction of a progression model is straightforward. For instance, to use CAPRI algorithm - with its default settings - and set a few graphical parameters to scale nodes size as of events' frequencies, scale legend which gets displayed in top corners and show p-values for the two elements which constitute the statistical evidence of selective advantage (temporal priority and probability raising) and hypergeometric test.

```
> tronco.plot(
>     tronco.capri(tronco.data),
>     scale.nodes = .6,
>     legend.cex = .6,
>     legend.pos = 'top',
>     confidence = c('tp', 'pr', 'hg'))
```

In a similar way, one might download, e.g., GISTIC data for CNAs, merge it to these mutations - via TRONCO's functions - and build a model with both data types included.

## 2.2 Clonal architecture of Clear Cell Renal Cell Carcinoma (CCRCC)

We now adopt the public data from *Gerlinger et al., Nat Gen 46, 2014* to show that the presented algorithms can be used to successfully reconstruct the clonal architecture in individual patients. We consider data coming from multi-region targeted exome-sequencing with high coverage ($> 70x$), which are reported in the reference paper as figures and in the supplementary materials.

Text files of the genetic profiles provided in such supplementary resources have been given as input to TRONCO and imported with the following R function (i.e., `import.sample`).

```r
# Import data - this converts the data from the CSV text file
# obtained from Excel, to a TRONCO object
import.sample = function(filename) {
        data = as.matrix(read.csv(paste0(filename), header=F))
        curr.data = NULL
        curr.data$genotypes = apply(data[c(-1,-2), -1, drop=F],
                2, as.numeric)
        rownames(curr.data$genotypes) =  data[c(-1,-2),1]
        curr.data$annotations = matrix(c(data[2, -1], data[1, -1]),
                ncol=2)
        event.names = c()
        for(i in 1:nrow(curr.data$annotations)){
                event.names = c(event.names, paste0("G",i))
        }
        colnames(curr.data$genotypes) = event.names
        rownames(curr.data$annotations) = event.names
        colnames(curr.data$annotations) = c("type", "event")
        curr.data$types = matrix('', nrow=length(unique(data[2, -1])),
                ncol=1)
        colnames(curr.data$types) = "color"
        rownames(curr.data$types) = unique(data[2, -1])
        for(i in 1:nrow(curr.data$types)) {
                curr.type = rownames(curr.data$types)[i]
                index = which(types == curr.type)
                curr.data$types[i,1] = colors[index]
        }
        print(curr.data)
        return(curr.data)
}
```

As an example, we next show the R code used to import the data for one of the patients ($RMH004$) from the study.

```
> library("TRONCO")
> library(RColorBrewer)
> colors = c(brewer.pal(6, "Set2"), "firebrick4", "dodgerblue4")
> types = c("Merged", "Frame Shift", "Splice site", "SNV",
> "Stop codon", "Disrupts start codon", "Gain",
> "Loss")
> rmh004.mut = import.sample("RMH004.csv")
$genotypes
    G1 G2 G3 G4 G5 G6
R3   1  1  0  0  0  0
VT   1  0  0  1  0  0
R10  1  0  0  0  1  0
R4   1  0  1  0  1  0
R2   1  0  1  0  1  1
R8   1  0  1  0  1  0
X    0  0  0  0  0  0

$annotations
   type           event
G1 "Frame Shift"  "VHL"
G2 "SNV"          "SMARCA4"
G3 "Merged"       "PBRM1(FS)/ATM(SC)"
G4 "Merged"       "PBRM1(FS)/ARID1A(SNV)"
G5 "Frame Shift"  "PTEN"
G6 "Merged"       "MSH6"

$types
              color
Frame Shift   "#FC8D62"
SNV           "#E78AC3"
Merged        "#66C2A5"
```

Now we load the CNAs and trim any event with no observations

```
> rmh004.CNA = import.sample("RMH004G.csv")
$genotypes
    G1 G2 G3 G4 G5 G6 G7 G8 G9 G10 G11
VT   0  0  1  0  0  0  0  0  1  0   0
R2   1  0  1  1  0  0  1  0  1  1   0
R3   0  0  1  0  0  0  0  0  1  0   0
R4   1  0  1  1  0  0  1  0  1  1   1
R8   1  1  1  1  0  1  1  0  1  1   0
R10  0  0  1  0  0  0  0  0  1  0   0
X    0  0  0  0  0  0  0  0  0  0   0

$annotations
   type    event
```

```
G1  "Gain"   "1q25.1"
G2  "Gain"   "2q14.3"
G3  "Gain"   "5q35.3"
G4  "Gain"   "7q22.3"
G5  "Gain"   "8q24.21"
G6  "Gain"   "12p11.21"
G7  "Gain"   "20q13.33"
G8  "Loss"   "1p36.11"
G9  "Loss"   "3p25.3"
G10 "Merged" "4q34.3-/8p23.2-/9p21.3-/14q31.1-"
G11 "Loss"   "6q22.33"

$types
        color
Gain    "firebrick4"
Loss    "dodgerblue4"
Merged  "#66C2A5"

> rmh004.CNA = trim(rmh004.CNA)
```

We use the `intersect` function of TRONCO to merge mutations profile with CNAs

```
> rmh004.data = intersect.datasets(rmh004.CNA, rmh004.mut,
> intersect.genomes = F)
*** Binding events for 2 datasets.
*** Intersect dataset [ intersect.genomes = FALSE ]
        x y result
Samples 7 7      7
Genes   9 6     15
```

The obtained TRONCO object has then been further edited in order to detect and handle genes with the same mutational profiles. We have to process these data as some alterations have the same genomic profile e.g., 5q35.3 amp, VHL fs and 3p25.3 loss. We get a report of these events with the function `consolidate.data`.

```
> duplicated.events = consolidate.data(rmh004.data, T)

Indistinguishable events:
    type      event
G1  "Gain"    "1q25.1"
G4  "Gain"    "7q22.3"
G6  "Gain"    "20q13.33"
G8  "Merged"  "4q34.3-/8p23.2-/9p21.3-/14q31.1-"
G12 "Merged"  "PBRM1(FS)/ATM(SC)"
Total number of events for these genes: 5
```

```
Indistinguishable events:
    type    event
G2  "Gain"  "2q14.3"
G5  "Gain"  "12p11.21"
Total number of events for these genes: 2

Indistinguishable events:
     type           event
G3  "Gain"         "5q35.3"
G7  "Loss"         "3p25.3"
G10 "Frame Shift"  "VHL"
Total number of events for these genes: 3
```

We use `merge.events` to merge the indistinguishable events

```
> rmh004.data = merge.events(rmh004.data,
> "G1", "G4", "G6", "G8", "G12",
> new.event = '1q 7q 20q 4q .. PBRM1fs ATMsc',
> new.type = 'Merged',
> event.color = as.colors(rmh004.data)['Merged'])

*** Binding events for 2 datasets.
```

Then we re-check the IDs of the events as data were edited in the previous step

```
> duplicated.events = consolidate.data(rmh004.data, T)

Indistinguishable events:
    type    event
G1  "Gain"  "2q14.3"
G3  "Gain"  "12p11.21"
Total number of events for these genes: 2

Indistinguishable events:
     type           event
G2  "Gain"         "5q35.3"
G4  "Loss"         "3p25.3"
G6  "Frame Shift"  "VHL"
Total number of events for these genes: 3

> rmh004.data = merge.events(rmh004.data,
> "G1", "G3",
> new.event = '2q 12p',
> new.type = 'Merged',
> event.color = as.colors(rmh004.data)['Merged'])
```

```
*** Binding events for 2 datasets.

> duplicated.events = consolidate.data(rmh004.data, T)

Indistinguishable events:
   type          event
G1 "Gain"        "5q35.3"
G2 "Loss"        "3p25.3"
G4 "Frame Shift" "VHL"
Total number of events for these genes: 3

> rmh004.data = merge.events(rmh004.data,
> "G1", "G2", "G4",
> new.event = '5q 3p VHL fs',
> new.type = 'Merged',
> event.color = as.colors(rmh004.data)['Merged'])

*** Binding events for 2 datasets.
```

The data are not yet ready for the subsequent analysis

```
> consolidate.data(rmh004.data, T)

$indistinguishable
list()

$zeroes
list()

$ones
list()
```

Besides quantification of intra-tumor heterogeneity, in their work *Gerlinger et al.* found that loss of the 3p arm and alterations of the Von Hippel-Lindau tumor suppressor gene *VHL* are the only events ubiquitous among their patients.

In Figure 2 we show the clonal evolution estimated for one of those patients, $RMH004$, computed with CAPRESE (shrinkage coefficient $\lambda = 0.5$, time $< 1$ sec) from the Bernoulli 0/1 profiles provided in Supplementary Table 3 and Figure 4 of the reference paper, with non-parametric bootstrap confidence (time $< 1$ sec).

```
> rmh004.data = tronco.caprese(rmh004.data)
*** Checking input events.
*** Inferring a progression model with the following settings.
        Dataset size: n = 7, m = 8.
        Algorithm: CAPRESE with shrinkage coefficient: 0.5.
```

```
The reconstruction has been successfully completed in 00h:00m:00s

> rmh004.data = tronco.bootstrap(rmh004.data)
Executing now the bootstrap procedure, this may take a long time...
Expected completion in approx. 00h:00m:00s

Performed non-parametric bootstrap with 100 resampling and 0.5
as shrinkage parameter.
```

This model may be compared to the one inferred by processing the region-specific VAF with a max-mini optimization of most parsimonious evolutionary trees, and performing selection-by-consensus when multiple optimal solutions exist - Supplementary Figure 9 of the reference paper. CAPRESE requires no arbitrarily defined curation criteria to select the optimal tree, as it constructively searches for a solution which, in this case, is analogous in suggesting *parallel evolution of subclones* via deregulation of the $SWI/SNF$ chromatin-remodeling complex – i.e., as may be noted from multiple clones with distinct *PBMR1* mutations.

Moreover, as a further proof of concept, we now consider the somatic mutations made available for two more patients, i.e., $EV002$ and $EV007$, once again as Bernoulli 0/1 profiles provided in Supplementary Table 3 and Figure 4 of the main study, of which the oncoprints follow in figure 3.

```
> oncoprint(example1.ev002,
> title="Phylogeny inference -
> Multiple biopses of patient EV002 (Gerlinger at al.)")
*** Oncoprint for "Phylogeny inference -
        Multiple biopses of patient EV002 (Gerlinger at al.)"
with attributes: stage=FALSE, hits=TRUE
Sorting samples ordering to enhance exclusivity patterns.
Setting automatic row font (exponential scaling): 13.6

> oncoprint(example2.ev007,
> title="Phylogeny inference -
> Multiple biopses of patient EV007 (Gerlinger at al.)")
*** Oncoprint for "Phylogeny inference -
        Multiple biopses of patient EV007 (Gerlinger at al.)"
with attributes: stage=FALSE, hits=TRUE
Sorting samples ordering to enhance exclusivity patterns.
Setting automatic row font (exponential scaling): 13.8
```

The inference on such patients is performed again with CAPRESE algorithm with shrinkage coefficient 0.5 (optimal value), and requires less than 1 second in a standard laptop; non-parametric bootstrap confidence is subsequently estimated.

We finally show the reconstructions (figure 4) of the above patients and we observe that once again CAPRESE captures the same clonal evolution depicted

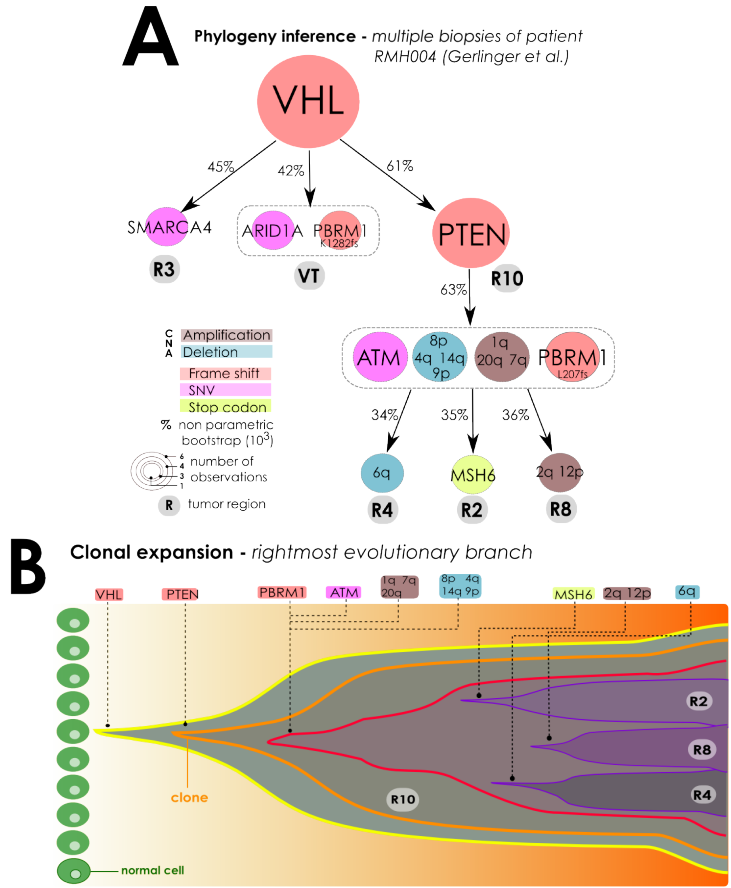

Figure 2: **(A)** With data provided by Gerlinger *et al.*, we infer a patient-specific clonal evolution from 6 biopsies of a clear cell renal carcinoma (5 primary tumor, 1 from the thrombus in the renal vein, *VT*). Validated non-synonymous mutations are selected for *VHL*, *SMARCA4*, *PTEN*, *PBMR1*, *ARID1A*, *ATM* and *MSH6* genes. CNAs are detected on 12 chromosomes. For this patient, both region-specific allele frequencies and Bernoulli profiles are provided. Thus, we can extract a clonal tree, signature and diffusion of each clone, by the unsupervised CAPRESE algorithm. **(B)** The unsupervised model inferred by CAPRESE predicts an analogous clonal expansion observed in the main paper, and extracted with most parsimonious phylogeny tree reconstruction from allelic frequencies, and hand-curated for selection of the optimal model. For simplicity, we show only expansion of the sub-clones harbouring *PTEN*'s frame shift mutation.

13

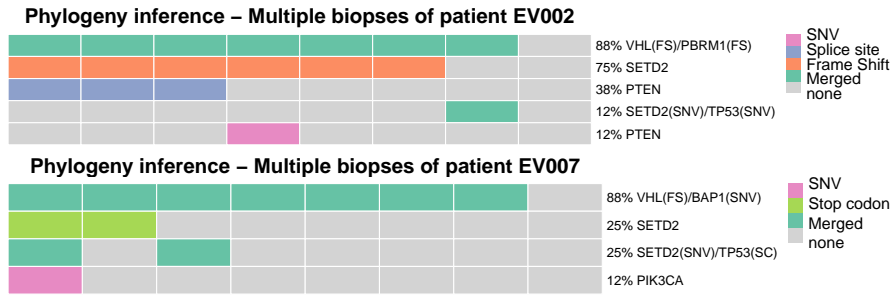

Figure 3: Oncoprint showing the genomic profiles of patients $EV002$ (top) and $EV007$ (bottom) from the study of *Gerlinger et al., Nat Gen 46, 2014*.

in the reference paper.

```
> tronco.plot(example1.ev002,scale.nodes=T,confidence=c("npb"),
> legend.pos="",label.edge.size=22,title="Inference by CAPRESE -
> Multiple biopses of patient EV002 (Gerlinger at al.)")
*** Expanding hypotheses syntax as graph nodes:
*** Rendering graphics
Nodes with no incoming/outgoing edges will not be displayed.
Set automatic fontsize scaling for node labels: 17.5622483502636
Adding confidence information: npb
RGraphviz object prepared.
Plotting graph and adding legends.

> tronco.plot(example2.ev007,scale.nodes=T,confidence=c("npb"),
> legend.pos="",label.edge.size=13,title="Inference by CAPRESE -
> Multiple biopses of patient EV007 (Gerlinger at al.)")
*** Expanding hypotheses syntax as graph nodes:
*** Rendering graphics
Nodes with no incoming/outgoing edges will not be displayed.
Set automatic fontsize scaling for node labels: 18.4548225555204
Adding confidence information: npb
RGraphviz object prepared.
Plotting graph and adding legends.
```

## 2.3 Atypical Chronic Myeloid Leukemia (aCML)

We now finally consider the data from *Piazza et al., Nat Gen 45, 2013* as a proof of usage of both CAPRESE and CAPRI algorithms on ensamble level data. Figure 5 shows the data used as inputs for both the two algorithms.

14

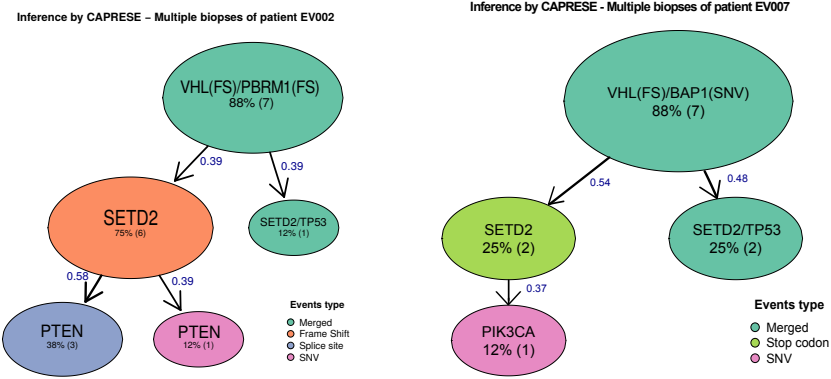

Figure 4: Reconstructions by CAPRESE for patients $EV002$ (left) and $EV007$ (right) from the study of *Gerlinger et al., Nat Gen 46, 2014.*

```
> require("TRONCO")
> data(aCML)
> aCML.alterations = events.selection(as.alterations(aCML),
+         filter.freq=.05, filter.in.names=c('KRAS', 'NRAS', 'IDH1',
+ 'IDH2', 'TET2', 'SF3B1', 'ASXL1'))
*** Aggregating events of type(s) {Ins/Del, Missense point, Nonsense
        Ins/Del, Nonsense point}
in a unique event with label "Alteration".
Dropping event types Ins/Del, Missense point, Nonsense Ins/Del,
        Nonsense point for 23 genes.

*** Binding events for 2 datasets.
*** Events selection: #events=23, #types=1 Filters freq|in|out =
        {TRUE, TRUE, FALSE}
Minimum event frequency: 0.05 (3 alterations out of 64 samples).

Selected 7 events.

[filter.in] Genes hold: KRAS, NRAS, IDH1, IDH2, TET2 ... [6/7 found].
Selected 10 events, returning.

> aCML.alterations = change.color(aCML.alterations,
> type="Alteration", new.color="khaki4")
> oncoprint(aCML.alterations)
*** Oncoprint for ""
with attributes: stage=FALSE, hits=TRUE
Sorting samples ordering to enhance exclusivity patterns.
Setting automatic row font (exponential scaling): 12.3
```

Moreover, we gave as a further input to CAPRI algorithm patterns of mutual exclusivity among genes: *TET2* and *IDH2*, *ASXL1* and *SF3B1*, *NRAS* and *KRAS*.

```
> aCML.alterations.hypo = hypothesis.add(aCML.alterations,
> 'TET2 xor IDH2', XOR('TET2','IDH2'))
> aCML.alterations.hypo = hypothesis.add(aCML.alterations.hypo,
> 'ASXL1 xor SF3B1', XOR('ASXL1','SF3B1'))
> aCML.alterations.hypo = hypothesis.add(aCML.alterations.hypo,
> 'NRAS xor KRAS', XOR('NRAS','KRAS'))
```

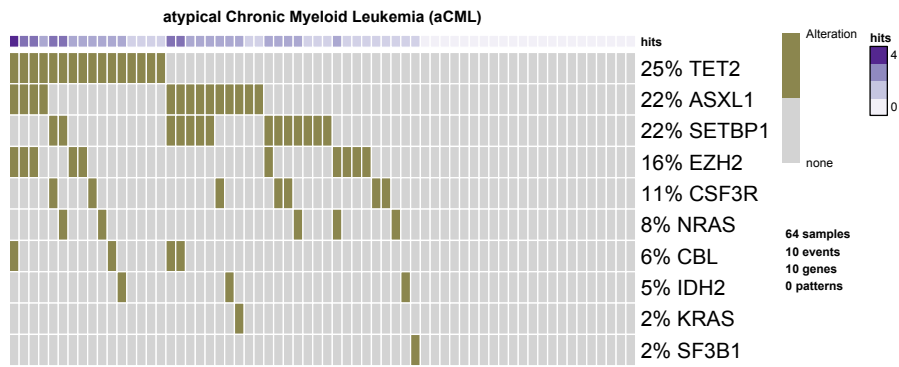

Figure 5: Oncoprint showing the genomic profiles of 64 aCML patients from the study of *Piazza et al., Nat Gen 45, 2013*.

We conclude by showing the results of the inference from both CAPRESE and CAPRI algorithms in figure 6.

```
> caprese = tronco.caprese(aCML.alterations)
*** Checking input events.
*** Inferring a progression model with the following settings.
        Dataset size: n = 64, m = 10.
        Algorithm: CAPRESE with shrinkage coefficient: 0.5.
The reconstruction has been successfully completed in 00h:00m:00s
> capri = tronco.capri(aCML.alterations.hypo)
*** Checking input events.
*** Inferring a progression model with the following settings.
        Dataset size: n = 64, m = 13.
        Algorithm: CAPRI with "bic, aic" regularization and
                  "hc" likelihood-fit strategy.
        Random seed: NULL.
        Bootstrap iterations (Wilcoxon): 100.
                  exhaustive bootstrap: TRUE.
```

```
                p-value: 0.05.
                minimum bootstrapped scores: 3.
*** Bootstraping selective advantage scores (prima facie).
        Evaluating "temporal priority" (Wilcoxon, p-value 0.05)
        Evaluating "probability raising" (Wilcoxon, p-value 0.05)
*** Loop detection found loops to break.
        Removed 5 edges out of 30 (17%)
*** Performing likelihood-fit with regularization bic.
*** Performing likelihood-fit with regularization aic.
The reconstruction has been successfully completed in 00h:00m:02s
There were 27 warnings (use warnings() to see them)
> tronco.plot(caprese, scale.nodes=F,
> title="aCML - Inference by CAPRESE")
*** Expanding hypotheses syntax as graph nodes:
*** Rendering graphics
Nodes with no incoming/outgoing edges will not be displayed.
Set automatic fontsize scaling for node labels: 15.2111016906551
Set automatic fontsize for edge labels: 7.60555084532756
Plotting graph and adding legends.
> tronco.plot(capri, scale.nodes=F,
> title="aCML - Inference by CAPRI")
*** Expanding hypotheses syntax as graph nodes:
*** Rendering graphics
Nodes with no incoming/outgoing edges will not be displayed.
Set automatic fontsize scaling for node labels: 16.2163594037787
Set automatic fontsize for edge labels: 8.10817970188937
Plotting graph and adding legends.
```

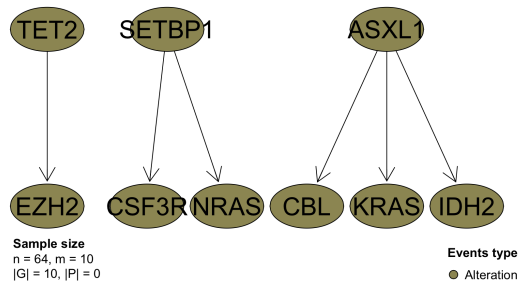
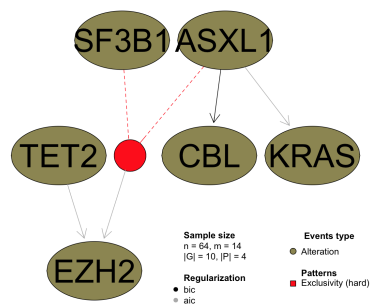

Figure 6: Inference by CAPRESE (left) and CAPRI (right) applied on the genomic profiles of 64 aCML patients from the study of *Piazza et al., Nat Gen 45, 2013*.

17



\end{document}